\documentclass[review,number,sort&compress]{elsarticle}
\usepackage{url}
\usepackage{amssymb}
\usepackage{amsmath}
\usepackage{mathtools}
\usepackage{multirow}
\usepackage{array}
\usepackage{subcaption}
\usepackage{comment}

\journal{***}

\begin{document}

\begin{frontmatter}

\title{Development of High-Sensitivity Radon Emanation Measurement Systems with Surface Treatment Optimization}

\author{Yuan Wu\fnref{fn1}}
%\author{Lin Si\corref{cor1}\corref{cor2}\fnref{fn1}}
\author{Lin Si\corref{cor1}\fnref{fn1}}
\author{Zhicheng Qian\fnref{fn1}}
\author{Youhui Yun\fnref{fn1}}
\author{Yue Meng \corref{cor1}\fnref{fn1,fn2,fn3}}
\author{Jianglai Liu\fnref{fn1,fn2,fn3,fn4}}
\author{Zhixing Gao\fnref{fn1}}
\author{Hao Wang\fnref{fn1}}
\author{Liangyu Wu\corref{cor2}\fnref{fn1}}
\author{Yuanzi Liang\fnref{fn5}}
\address[fn1]{School of Physics and Astronomy, Shanghai Jiao Tong University, Key Laboratory for Particle Astrophysics and Cosmology (MoE), Shanghai Key Laboratory for Particle Physics and Cosmology, Shanghai 200240, China}
\address[fn2]{Shanghai Jiao Tong University Sichuan Research Institute, Chengdu 610213, China}
\address[fn3]{Jinping Deep Underground Frontier Science and Dark Matter Key Laboratory of Sichuan Province}
\address[fn4]{New Cornerstone Science Laboratory, Tsung-Dao Lee Institute, Shanghai Jiao Tong University, Shanghai 201210, China}
\address[fn5]{College of Physics, Jilin University, Jilin 130012, China}
\cortext[cor1]{Corresponding author. Email: mengyue@sjtu.edu.cn (Yue Meng), silin123@sjtu.edu.cn (Lin Si, Now at Stanford University, Stanford, CA, USA)}
\cortext[cor2]{Now at Stanford University, Stanford, CA, USA}

\begin{abstract} Radon and its progenies are significant sources of background in rare event detection experiments, including dark matter searches like the PandaX-4T experiment and other rare decay studies such as neutrinoless double beta decay (NLDBD). In order to measure and control radon emanation for these experiments, we have developed two specialized radon measurement systems: a radon emanation measurement system suitable for small-sized samples with a blank rate of $0.03 \pm 0.01$ mBq in the 12.3 L counting chamber, and a radon trap system designed for large-volume samples using low-temperature radon trapping techniques, which improves the sensitivity by a factor of 30 with 1 standard liter per minute (slpm) gas flow and 6 hours trapping time. To boost the detection sensitivity, various surface treatments of the chambers were investigated, including mechanical polishing, electrochemical polishing, and mirror polishing, which reveals that smoother surfaces lead to lower radon emanation rates. In addition, treatments such as applying epoxy coating and covering with aluminized Mylar to stainless steel chambers can also reduce the radon emanation by ($90 \pm 7)\%$ and ($60 \pm 12)\%$, respectively.
\end{abstract}

\end{frontmatter}
%\linenumbers

\section{Introduction}
Rare event detection experiments, such as those searching for dark matter and NLDBD, demand extremely low levels of background radiation to achieve their scientific goals. Among the leading candidates for dark matter are Weakly Interacting Massive Particles (WIMPs), which have been driving extensive experimental efforts worldwide. Liquid xenon time projection chambers (TPCs), such as those employed in PandaX-4T, LUX-ZEPLIN (LZ), and XENON experiments, represent the state-of-the-art technique in WIMP detection due to their ability to achieve ultra-low background levels and exceptional sensitivity to rare interactions \cite{PandaX-4T:2021bab, PandaX:2024qfu, LZCollaboration:2024lux, XENON:2023cxc}. Beyond WIMP searches, such multi-purpose detectors also open avenues for complementary physics studies, such as the search for NLDBD, a rare nuclear process that would confirm the Majorana nature of neutrinos \cite{RevModPhys.87.137, annurev:/content/journals/10.1146/annurev-nucl-101918-023407, PandaX:2023ggs, PandaX:2022kwg}. Various experiments are focusing on detecting NLDBD in different isotopes. Notably, EXO-200/nEXO and KamLAND-Zen, focus on detecting NLDBD in xenon isotopes \cite{EXO-200:2019rkq, nEXO:2021ujk, KamLAND-Zen:2022tow}, while experiments such as GERDA and Majorana Demonstrator utilize germanium detectors \cite{GERDA:2020xhi, Majorana:2022udl}, and CUORE targets $^{130}$Te as a candidate isotope for NLDBD \cite{CUORE:2021mvw}.

A significant source of background in such experiments is radon, particularly $^{222}$Rn with a half-life of 3.82 days, which is continuously produced from the $^{238}$U decay chain and emanates from the detector materials. Among its decay products, $^{214}$Pb, which undergoes $\beta$-decay, is a significant contributor to electron recoil (ER) background for direct dark matter detection \cite{PandaX-4T:2021bab, PandaX-4T:2021lbm}, and it can also affect the sensitivity of NLDBD experiments by adding background noise in the signal energy range \cite{PandaX:2023ggs}. Therefore, precise measurement and control of radon-induced background are crucial. This requires an extensive screening campaign to measure the $^{222}$Rn emanation from detector materials in contact with detection target using a highly sensitive radon detector. 

Among various radon detection techniques, electrostatic collection is widely used due to its simplicity in installation and operation, capability for continuous measurements, and cost efficiency. This method collects positively charged radon progeny ($^{218}$Po and $^{214}$Po) onto a silicon detector as an alpha spectroscopy. Several large-scale experiments employ this technique: the Borexino radon monitor achieved a detection limit below 1 mBq/m$^3$ with a 418 L chamber \cite{KIKO2001272}; Super-Kamiokande developed an 80 L detector with a background of $0.33 \pm 0.07$ mBq/m$^3$ to monitor radon in its buffer gas \cite{Nakano:2017rsy}; JUNO experiment established a 41.5 L electrostatic collection chamber with a low background event rate of $0.70 \pm 0.15$ counts per day (cpd) \cite{Chen_2022}, and SuperNEMO's 50 L prototype detector demonstrated a sensitivity of 11 mBq/m$^3$ for one-day measurements \cite{Mamedov:2011zz}.

In this work, we have developed two radon measurement systems for rare decay experiments: a radon emanation measurement system for small samples and a radon trap system for large-volume samples. Additionally, surface treatment studies were conducted to minimize radon emanation from detection chamber materials, further enhancing sensitivity.

The paper is structured as follows: Section \ref{sec:Radon measurement system} discusses the configuration, design, layout, signal processing and sensitivity of radon emanation measurement system. Section \ref{sec:Radon trap system} presents the radon trap system, including its operating principle and trapping efficiency. Section \ref{sec:SurfaceTreatments} provides experimental data and analysis on the relationship between surface roughness and radon emanation and evaluates the performance of various surface treatments in reducing radon emanation. Section \ref{sec:Conclusion} is the conclusions.

\section{Radon emanation measurement system}\label{sec:Radon measurement system}

\subsection{Radon emanation detector setup}\label{subsec:radon emanation detector setup}
\begin{figure}[!th]
\begin{center}
\includegraphics[width=8cm]{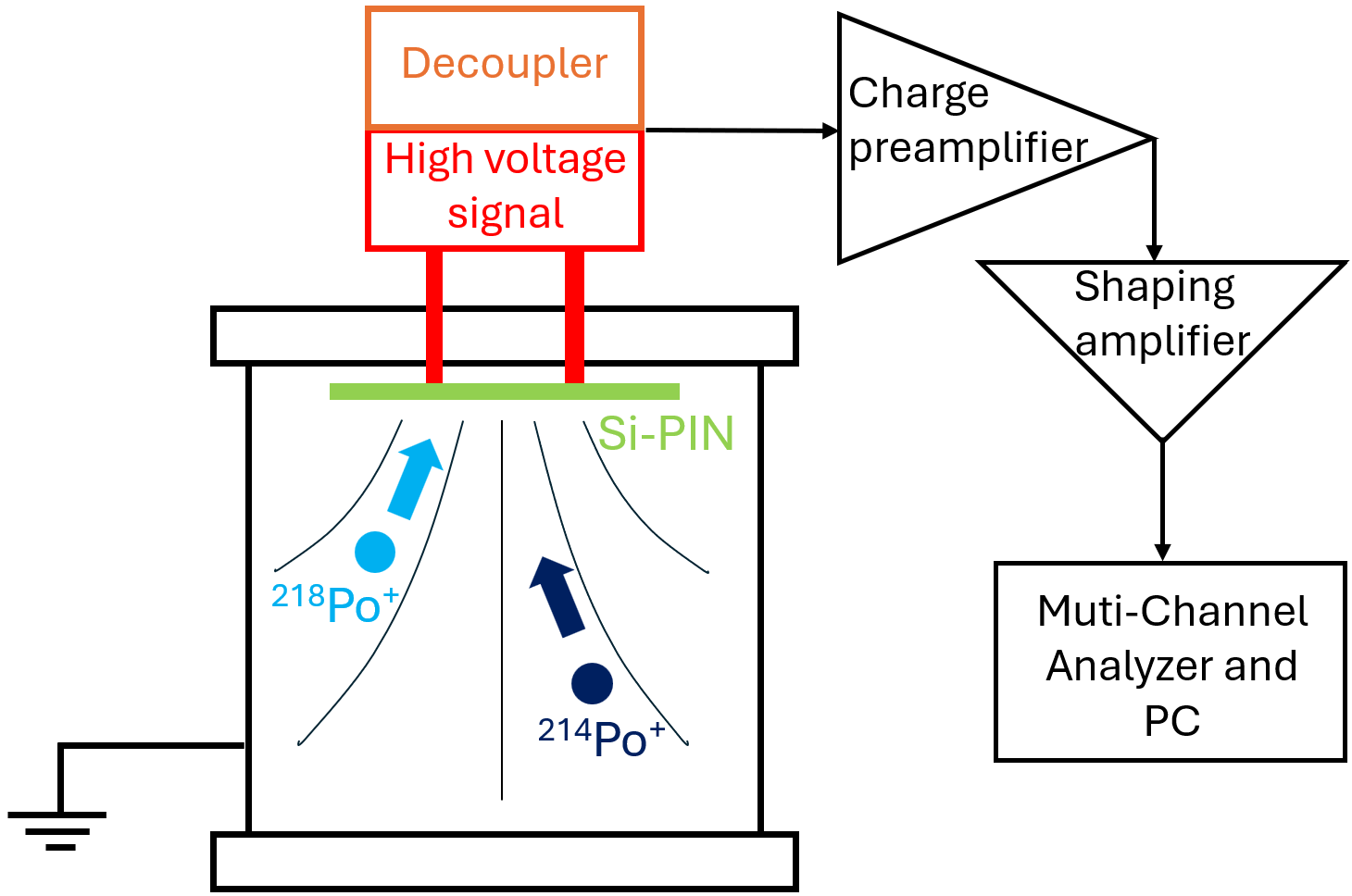}\\
\caption{A schematic view of the radon detector}
\label{fig:detector}
\end{center}
\end{figure}

The radon detector with the electrostatic method consists of a grounded cylindrical stainless steel chamber and a Si-PIN diode (Hamamatsu Photonics S3204-09 \cite{Si-PIN}). The Si-PIN is electrically isolated from the stainless steel vessel with two Safe High Voltage (SHV) coaxial feed-through connectors. A negative high voltage (HV) divider is constructed to provide an electric field with -1600 V voltage between the Si-PIN diode and the grounded chamber, forming a radial field that effectively collects positively charged radon daughters, such as $^{218}$Po and $^{214}$Po \cite{positive}, onto the surface of the Si-PIN. The collected radon progenies undergo alpha decays, which are detected by the Si-PIN. In addition, a bias voltage generated by the circuit board \cite{Mitsuda:2003eu} is applied to the Si-PIN diode to ensure its proper operation and signal readout. The energy deposited in the Si-PIN from alpha decays is converted into charge signals, which are processed by a charge amplifier (H4083 model) \cite{H4083} before being shaped and digitized by a digital multi-channel analyzer (MCA) (DT5780) \cite{DT5780} for digital nuclear spectroscopy. A schematic view of the detector setup is shown in Figure \ref{fig:detector}.

\begin{figure}[!th]
\begin{center}
\includegraphics[width=8cm]{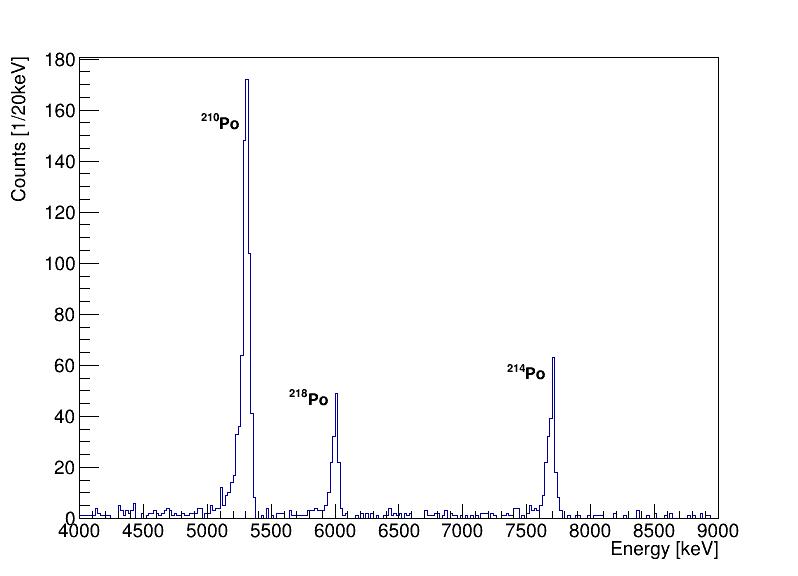}\\
\caption{A typical energy spectrum by the electrostatic collection detector. The three peaks, from left to right, correspond to $^{210}$Po (5.3 MeV), $^{218}$Po (6.0 MeV) and $^{214}$Po (7.7 MeV).}
\label{fig:spectrum}
\end{center}
\end{figure}

A typical energy spectrum obtained by the electrostatic collection detector is shown in Figure \ref{fig:spectrum}. The energy scale is calibrated in-situ using peaks from typical radon daughter alpha decays with source. The radon emanation rate is calculated using $^{214}$Po alpha decay (7.7 MeV) counting rate and detection efficiency because it offers a cleaner and more distinct energy peak. While $^{218}$Po alpha decay (6.0 MeV) energy range could be contaminated by alpha decays from $^{212}$Bi (6.1 MeV) and $^{210}$Po (5.3 MeV) \cite{nndc}, especially when noise levels are high, which can compromise the energy resolution. $^{212}$Bi, a decay product from the $^{232}$Th decay chain, can occasionally be present in measurements, but this contamination is typically less of a concern compared to the more prevalent $^{210}$Po contamination, which originates from the decay of $^{210}$Pb with a long half-life on the Si-PIN surface, and has been consistently detected over time. Additionally, we observed that when the Si-PIN diode is replaced, the $^{210}$Po activity changes, further confirming that the source of $^{210}$Po contamination originates from previous calibrations with radon source and sample measurements deposited on the Si-PIN surface.

Between $^{222}$Rn and $^{214}$Po, there are three intermediate isotopes: $^{218}$Po, $^{214}$Pb, and $^{214}$Bi, with half-lives of 3.1 min, 26.8 min, and 19.9 min, respectively. It takes a few hours to reach secular equilibrium for $^{214}$Po. During this initial period, data is discarded due to the non-equilibrium condition. Considering the 3.8-day half-life of radon, the typical monitoring periods are longer than ten days.

\subsection{System layout}\label{subsec:System design and layout}
\begin{figure}[!th]
\begin{center}
\includegraphics[width=8cm]{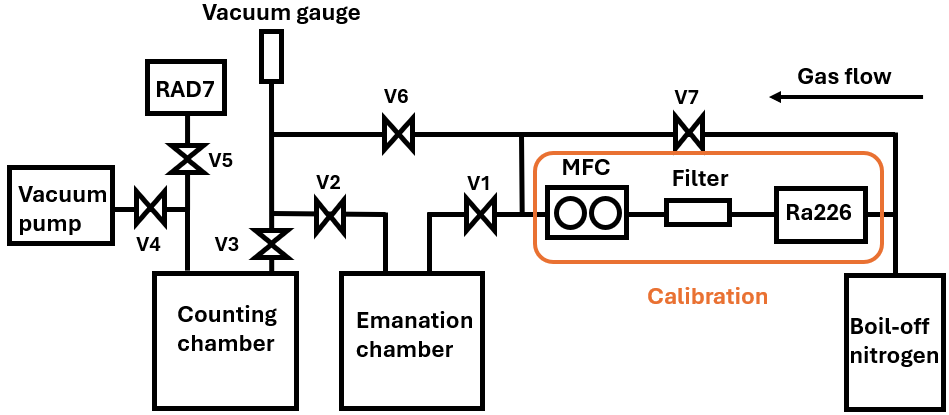}\\
\caption{A schematic diagram of the emanation measurement system}
\label{fig:emanation}
\end{center}
\end{figure}

The radon system is composed of a 12.33 L counting chamber and an emanation chamber with the same size. Figure \ref{fig:emanation} shows a schematic diagram of the system. The piping assemblies are 1/4 inch ultra-high vacuum (UHV) pipes and valves (HE Lennon Inc., Swagelok, Michigan) ~\cite{Swagelok}. To prevent water vapor from being absorbed into the chamber and cause detection efficiency loss \cite{Metal}, only boil-off nitrogen is used as carrier gas during the measurements.

At the beginning of each measurement, boil-off nitrogen flows through the entire system to purge water vapor and air radon out the chambers. The stop flushing time is documented, and valves 1 and 2 are closed. Then, we wait for at least ten days to allow the samples to emanate radon within the chamber. Subsequent to this emanation period, valves 3 and 4 are opened to activate the vacuum pump and evacuate the counting chamber. After evacuation, valve 4 is closed, and valve 2 is opened, allowing gas to transfer from the emanation chamber to the counting chamber until the pressure equilibrates, effectively transferring half of the radon atoms (when the volumes of the two chambers are equal). The valves are then closed, and radon counting commences at 25°C, under 4\% relative humidity.

To ensure accurate results, blanks measurements are performed to quantify potential contaminants or intrinsic radioactivity within the chambers. The blank of the emanation chamber quantifies radon emanation from its inner surface along with its connecting pipes and is denoted as \( A_0^{\text{blank}} \). Meanwhile, the blank of the counting chamber, denoted as \( R_C \), is derived from self-measurement without the emanation chamber and has been determined to be as low as \( 0.03 \pm 0.01 \) mBq, which was achieved by the mirror polishing method described in Section \ref{sec:SurfaceTreatments}. These blanks are measured under identical operational conditions as the actual experiments and are used to correct the measured sample activity, as described below.

\begin{figure}[!th]
\begin{center}
\includegraphics[width=8cm]{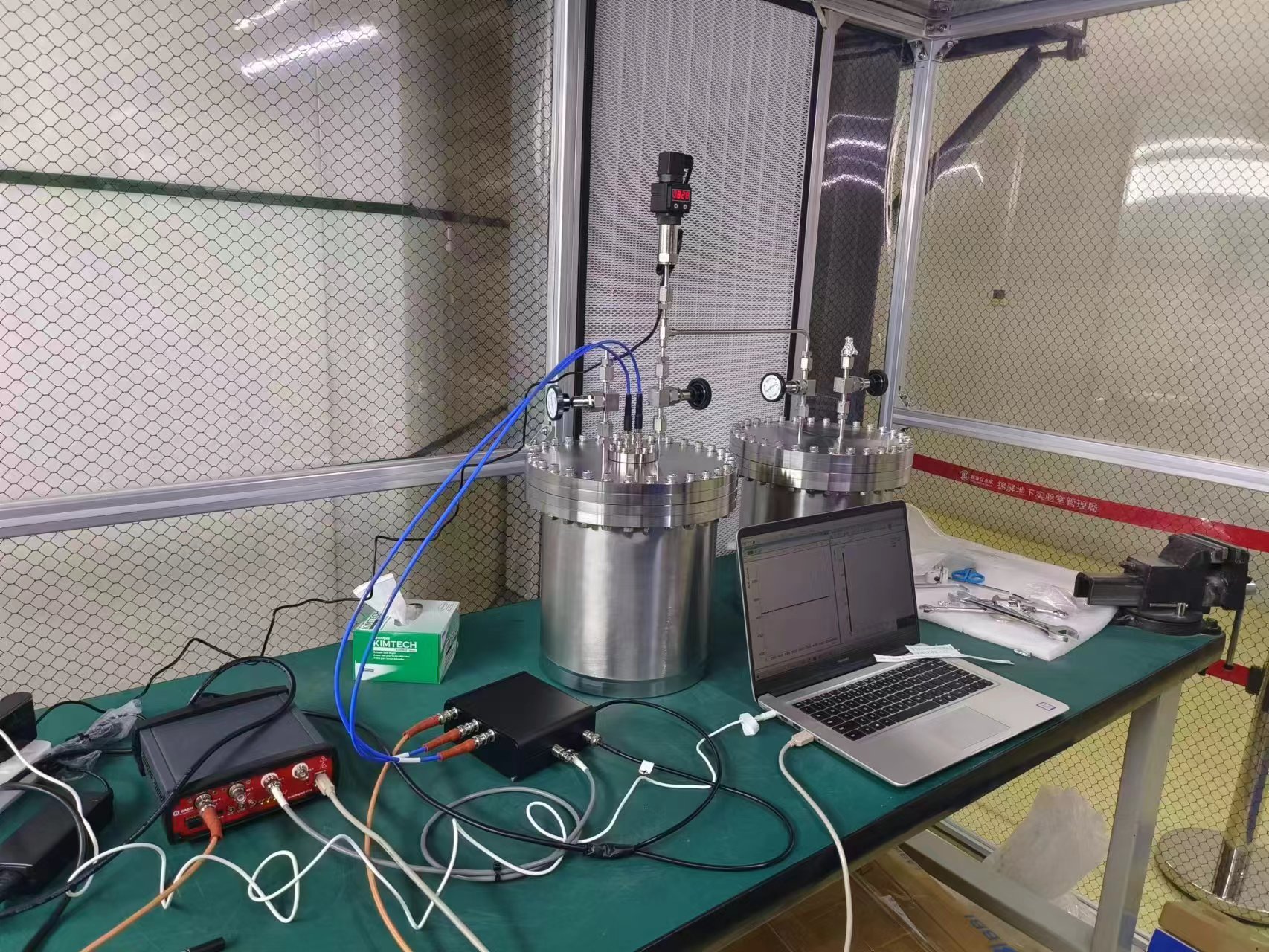}\\
\caption{A photo of the radon measurement system}
\label{fig:real_photo}
\end{center}
\end{figure}

In the counting chamber, radon decays over time, and additional radon emanates from the counting chamber walls at a rate \( R_C \). The number of radon atoms \( N(t) \) in the counting chamber at time \( t \) after the transfer satisfies the following equation:
\begin{equation}\label{eq:diff_eq}
\frac{dN(t)}{dt} = -\lambda N(t) + R_C,
\end{equation}
where,
\begin{itemize}
    \item \( N(t) \) is the number of radon atoms in the counting chamber at time \( t \),
    \item \( \lambda \) is the decay constant of radon,
    \item \( R_C \) is the radon emanation rate from the counting chamber (atoms per unit time).
\end{itemize}
Solving Equation~\eqref{eq:diff_eq} with the initial condition \( N(0) = N_0 \), where \( N_0 \) is the number of radon atoms transferred from the emanation chamber, we obtain:
\begin{equation}\label{eq:N_t}
N(t) = \frac{R_C}{\lambda} + \left( N_0 - \frac{R_C}{\lambda} \right) e^{-\lambda t}.
\end{equation}
The activity \( A(t) \) in the counting chamber at time \( t \) is given by:
\begin{equation}\label{eq:A_t}
A(t) = \lambda N(t) = R_C + \left( A_0 - R_C \right) e^{-\lambda t},
\end{equation}
where \( A_0 = \lambda N_0 \) is the initial activity due to radon transferred from the emanation chamber.

\begin{figure}[!th]
\begin{center}
\includegraphics[width=8cm]{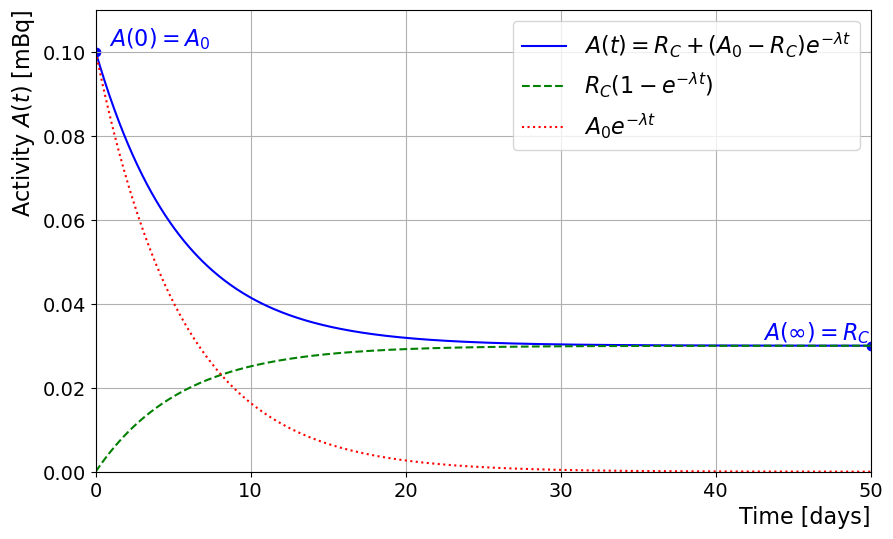}\\
\caption{Radon activity \( A(t) \) in the counting chamber over time, assume \( 0.1 \) mBq radon transferred from emanation chamber.}
\label{fig:A_t_plot}
\end{center}
\end{figure}

This equation shows that the measured activity consists of a decaying component from the radon transferred (from both the sample and the emanation chamber) and a growing component from radon emanation of the counting chamber itself as illustrated in Figure \ref{fig:A_t_plot}. At \( t = 0 \), the activity is \( A(0) = A_0 \), and as \( t \rightarrow \infty \), it approaches \( A(\infty) = R_C \). This reflects the fact that initially, the activity is dominated by radon transferred from the emanation chamber, and over time, as this radon decays, the activity approaches the background level due to radon emanation from the counting chamber.

To determine the radon emanation rate \( R_S \) from the sample, we measure the activity \( A(t) \) in the counting chamber over time after the transfer, and fit the data to Equation~\eqref{eq:A_t} to extract the initial activity \( A_0 \) and the background rate \( R_C \). We then compute the net sample activity by correcting \( A_0 \) for the total system efficiency \( \epsilon_{\text{total}} \), and subtracting the contribution from a blank measurement without the sample (\( A_0^{\text{blank}} \)):
\begin{equation}\label{eq:A_sample}
A_{\text{sample}} = \frac{A_0 - A_0^{\text{blank}}}{\epsilon_{\text{total}}}.
\end{equation}

\subsection{Efficiency investigation}\label{subsec:eff-investigation}

To measure the sample, two modes are employed depending on the condition: the static mode, as described in Section \ref{subsec:System design and layout}, directly measures radon emanation from a sample placed in the chamber without continuous gas flow; the dynamic mode, on the other hand, is used when the system being measured is in gas circulation with the detector, enabling continuous monitoring. To calibrate the efficiency of the electrostatic detector in the two modes, a $^{226}$Ra source (“flow-through" $^{226}$Ra source from Nanhua University) with known activity was used \cite{Takenaka:2024ctk}, which provides a radon concentration of about 200 Bq/m$^3$ with a 1 slpm N$_2$ flush. In the dynamic mode, gas flow continuously through the source to the detector. In contrast, the static mode measures decay activity after introducing a set amount of radon into the chamber.

The calibration setup involved connecting the flow meter control, sources, detector, and a commercial radon detector RAD7 \cite{durridge_rad7} in sequence. Calibration was initiated in the dynamic mode with nitrogen gas at a flow rate of 1 slpm, with radon concentration monitored in real-time by RAD7. After the radon concentration stabilized, the valve was closed to transition to the static mode. In this static phase, the decay of the known amount of radon within the detector was measured. By fitting the decay curve, the initial $^{214}$Po signal rate was deduced. For both dynamic and static modes, the efficiency of the detector $\epsilon_{\text{counting}}$ is determined using the following equation:
\begin{equation}\label{equ:efficiency}
\epsilon_{\text{counting}} = \frac{R_{\text{$^{214}$Po}}}{A_{\text{$^{222}$Rn}}}
\end{equation}
In this equation, the numerator represents the $^{214}$Po signal rate detected by the Si-PIN photodiode (counts per second), and the denominator is the $^{222}$Rn activity (Bq) of the source, which can be measured by the RAD7.

\begin{figure}[!th]
\begin{center}
\includegraphics[width=8cm]{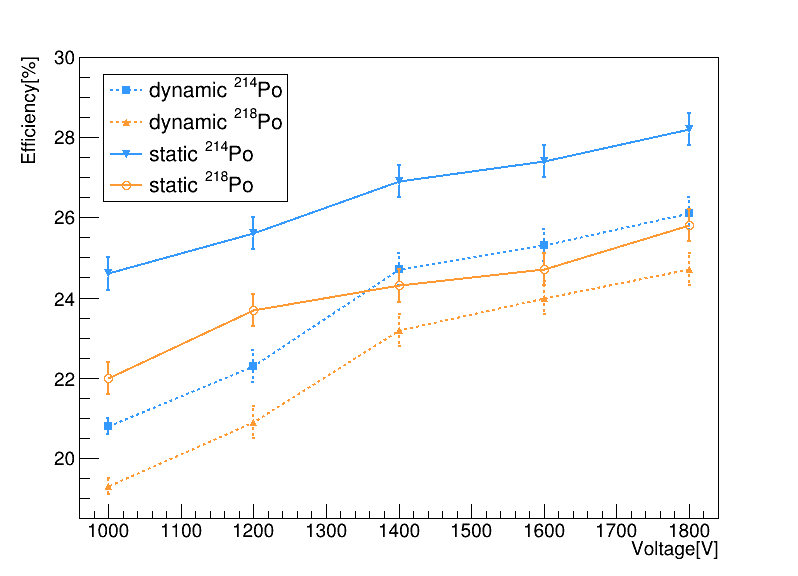}\\
\caption{The collection efficiencies of $^{214}$Po and $^{218}$Po at different high voltages. Measurements include dynamic mode and static mode for both isotopes.}
\label{fig:eff}
\end{center}
\end{figure}

We further explored the impact of applied high voltage on the collection efficiencies of $^{214}$Po and $^{218}$Po under the dynamic mode and the static mode, as illustrated in Figure \ref{fig:eff}. Measurements were conducted at applied high voltages ranging from \(-1000\) V to \(-1800\) V. The result shows that the efficiencies for both isotopes increased with higher applied voltage, indicating better collection at higher electric field strength. The static mode provides better efficiencies due to a more stable collection environment, while the dynamic mode shows lower efficiencies, likely because gas movement disturbs ion collection. The $^{214}$Po efficiency is higher than that of $^{218}$Po, primarily because the collected $^{218}$Po will continue to decay into $^{214}$Po on the PIN diode, and the uncollected $^{218}$Po will decay to intermediate isotopes $^{214}$Pb and $^{214}$Bi, which are also positively charged, can be collected onto the PIN diode. These isotopes will decay into $^{214}$Po, adding to the $^{214}$Po signal. Moreover, $^{218}$Po may be more susceptible to neutralization, particularly in high humidity environments. When $^{218}$Po combines with water molecules, it loses its charge and becomes unable to be collected by the electrostatic field, whereas $^{214}$Po seems to be less affected by this phenomenon, which could help explain why its efficiency is higher.

The entire system's efficiency, which includes the emanation chamber, is given by:
\begin{equation}\label{equ:scale-factor}
\epsilon_{\text{total}} = \epsilon_{\text{counting}} \times \frac{V_{\text{counting}}}{V_{\text{counting}} + V_{\text{emanation}}},
\end{equation}
where $V_{\text{counting}}$ and $V_{\text{emanation}}$ represent the volumes of the counting and emanation chambers, respectively.

\section{Radon trap system}\label{sec:Radon trap system}
To enhance the sensitivity of the radon emanation rate, a radon trap system is introduced. This system operates by cooling the radon carrier gas to liquid nitrogen temperature, where radon transitions from the gaseous to the solid state due to its relatively high boiling point. The radon is then trapped in the cold trap, enabling more efficient collection and detection.
\subsection{Layout}
\begin{figure}[!th]
\begin{center}
\includegraphics[width=8cm]{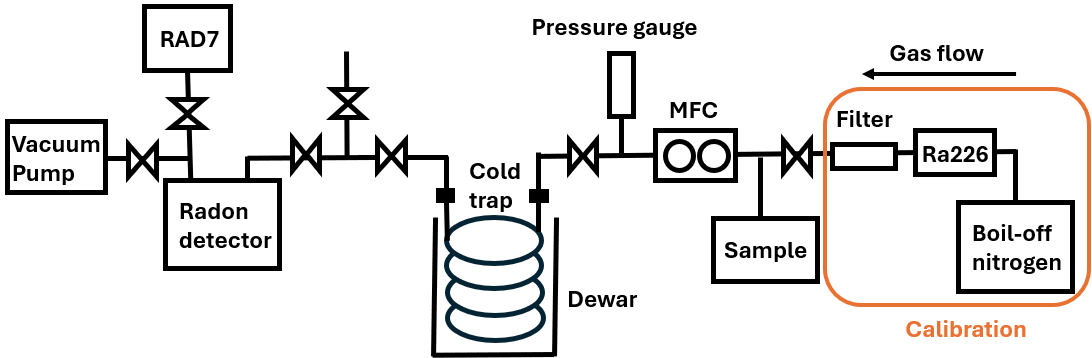}\\
\caption{A schematic diagram of a radon trap system}
\label{fig:coldtrap}
\end{center}
\end{figure}

As described in Figure \ref{fig:coldtrap}, the radon trap system consists of a cold trap, a flow control meter, a pressure gauge, a radon detector and a vacuum pump.The system should have 100\% metal-seals such as VCR\textsuperscript{\textregistered} metal gasket face seal \cite{Swagelok} or Conflat Flange (CF). The cold trap itself is a 1/4 “spiral" Ultra High Purity (UHP) stainless steel tube that is suitable for submersion in a liquid nitrogen bath.

\subsection{Trapping efficiency}
To determine the trapping efficiency of the radon trap system, the $^{226}$Ra source, which provides a high concentration of radon, is flushed with a carrier gas for approximately fourteen hours to stabilize the radon level within the system. During this flushing period, radon passes through the cold trap, where a mass flow meter regulates the gas flow rate. The pressure within the system is continuously monitored via a pressure gauge. Additionally, a RAD7 detector is integrated into the setup, as described in Section \ref{subsec:eff-investigation}, to continuously monitor the radon concentration and ensure it reaches a stable value.

Once the radon concentration is stabilized, the cold trap is immersed in liquid nitrogen, and a reference time is established. After the predefined trapping period, the valves at both ends of the cold trap are sealed, the liquid nitrogen is removed, and the system is allowed to return to room temperature. At the same time, the detector is evacuated, and the valve between the cold trap and the detector is opened, allowing the radon gas to enter the detector due to the pressure differential. The amount of radon that enters the detector ($A_1$) is then determined by monitoring the decay rate using our detection setup.

The efficiency of radon trapping ($\epsilon_{trap}$) is defined as the ratio of $A_1$ to $A_0$:
\begin{equation}\label{eq:trapEfficiency}
    \epsilon_{trap} = \frac{A_1}{A_0} = \frac{A_1}{C \cdot F \cdot \Delta{t}}
\end{equation}
where \(A_0\) represents the total radon amount (Bq), calculated as the product of the radon concentration \(C\) measured by RAD7 (Bq/m\(^3\)), flow rate \(F\) (slpm), and duration \(\Delta{t}\) (min), while \(A_1\) is the activity measured by our detector, adjusted for the detection efficiency.

The overall efficiency of the system is a combination of the detection efficiency ($\epsilon_{counting}$) and the cold trap efficiency ($\epsilon_{trap}$), as defined by Equations \ref{equ:efficiency} and \ref{eq:trapEfficiency}.

According to our measurements, using the aforementioned cold trap enrichment method with boil-off nitrogen gas at 1 slpm through the radon source for 6 hours, the trapping efficiency is (89 ± 4)\%. Based on the amount of gas trapped and the trapping efficiency, the radon trap system will boost the sensitivity by approximately 30 times under the described conditions, making it particularly effective for large-volume containers measurement.

\section{Surface treatments on stainless steel}\label{sec:SurfaceTreatments}

Surface treatments on stainless steel can significantly affect radon emanation rates. This section explores the impact of different treatments on the radon release rates, focusing on surface polishing and other treatments such as epoxy coating and aluminized Mylar membrane covering.

\subsection{Relationship between surface roughness and radon emanation}\label{sec:Relationship between roughness and radon release}

Our study utilized two radon detectors with volumes of 7.4 L and 12.3 L, respectively, to measure the radon emanation rates from their inner surfaces with varying surface roughness, achieved using different polishing methods. The experimental setup was designed to ensure controlled conditions, where the only variable was the roughness of the inner surfaces, measured by surface roughness tester TIME\textsuperscript{\textregistered} 3200 \cite{TIME3200}.

\begin{itemize}
    \item Mechanical Polishing:  
        Mechanical polishing was performed using abrasive materials with progressively finer grits to remove surface imperfections and achieve a smoother surface. This process began with rough abrasives, followed by finer abrasives to reduce surface roughness. The final polishing stages used a cloth wheel to further enhance smoothness.
    
    \item Electrochemical Polishing:  
        Electrochemical polishing was carried out by immersing the stainless steel samples in an electrolyte solution. The samples acted as the anode, and when an electric current was applied, material was dissolved from the surface, smoothing out microscopic peaks and valleys. This process not only improved surface roughness but also enhanced corrosion resistance and cleanliness.
    
    \item Mirror Polishing:  
        Mirror polishing involved the use of finer abrasives and polishing compounds to achieve a highly reflective surface. Polishing waxes were used during the final stages of the process. A wool or cloth wheel, driven by a high-speed motor, was coated with the polishing wax and applied to the stainless steel surface to remove micro-scratches and achieve a mirror-like finish. This method achieved the lowest surface roughness values and the highest reduction in radon emanation rates.
\end{itemize}

After polishing process, the chambers underwent ultrasonic cleaning with pure water, followed by wiping with alcohol and acetone to ensure the removal of any residual polishing compounds. 

\begin{figure}[!th]
\begin{center}
\includegraphics[width=8cm]{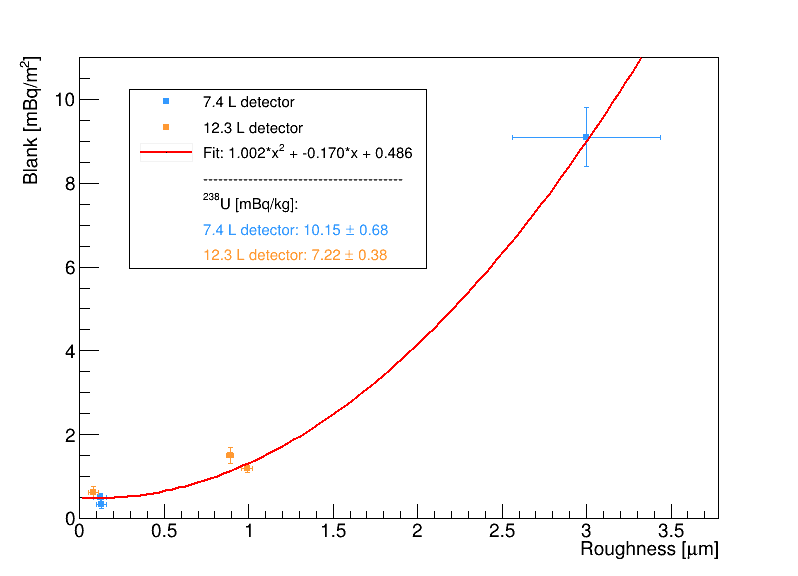}\\
\caption{Relationship between roughness and radon emanation, with intrinsic $^{238}$U radioactivity of the two detectors}
\label{fig:roughness}
\end{center}
\end{figure}

The radon emanation rates were plotted against the roughness of the chambers’ inner surface to identify any potential trends. The data indicated a clear relationship, as shown in Figure \ref{fig:roughness}, where an increase in roughness correlates with higher radon emanation rates.

A second-degree polynomial function was employed to model the relationship between surface roughness and radon emanation rates. It indicates that surface roughness is a significant predictor of radon release for stainless steel.

The observed trend can be attributed to the increased surface area and the presence of micro-crevices in rougher surfaces, which provide more sites for radon progeny to form and subsequently escape. The dominant mechanism for radon release is alpha recoil, where radon atoms are ejected from the surface due to the energy released during radioactive decay \cite{Jorg:2022spz}. On rough surfaces, these  irregular shapes and crevices provide more angles and pathways for the radon atoms to escape rather than being reabsorbed by the material.

In addition to radon emanation rates, small pieces cut from the stainless steel chambers were measured for $^{238}$U radioactivity using NexION 5000 Multi-Quadrupole ICP Mass Spectrometer \cite{PandaX-4T:2021lbm, perkinelmer2023}. Higher $^{238}$U radioactivity in stainless steel generally correlates with higher radon emanation rates since radon is a decay product of $^{238}$U. However, in this study, the $^{238}$U concentrations of the two chambers were similar, resulting in comparable radon emanation rates. This suggests that, under the conditions of this experiment, surface roughness, rather than $^{238}$U radioactivity, is the primary factor influencing the radon emanation. The roughness values achieved using different polishing methods for the inner surface of stainless steel chambers, along with their corresponding radon blanks and $^{238}$U concentrations, are summarized in Table \ref{tab:RoughnessPolishing}.

\begin{table}[h!]
\centering
\caption{Performance of various polishing methods on radon blanks and $^{238}$U concentrations.}
\label{tab:RoughnessPolishing}
\small
\begin{tabular}{|m{2.6cm}|m{2.6cm}|m{2.75cm}|m{2.65cm}|}
\hline
\textbf{Polishing} & \textbf{Roughness [um]} & \textbf{Blank [mBq/m$^2$]} & \textbf{$^{238}$U [mBq/kg]} \\
\hline
\multicolumn{4}{|c|}{7.4 L SS Detector (Surface Area: 0.21 m$^2$)} \\
\hline
Electrochemical & \centering 3.00 ± 0.44 & \centering 9.08 ± 0.71 & \multirow{3}{=}{\centering 10.15 ± 0.68} \\
\cline{1-3}
Mirror Polishing & \centering 0.12 ± 0.04 & \centering 0.48 ± 0.14 & \\
\cline{1-3}
Mirror Polishing + Electrochemical & \centering 0.13 ± 0.03 & \centering 0.33 ± 0.10 & \\
\hline
\multicolumn{4}{|c|}{12.3 L SS Detector (Surface Area: 0.30 m$^2$)} \\
\hline
Mechanical & \centering 0.99 ± 0.03 & \centering 1.18 ± 0.14 & \multirow{3}{=}{\centering 7.22 ± 0.38} \\
\cline{1-3}
Mechanical + Electrochemical & \centering 0.89 ± 0.02 & \centering 1.52 ± 0.20 & \\
\cline{1-3}
Mirror Polishing & \centering 0.08 ± 0.03 & \centering 0.64 ± 0.14 & \\
\hline
\end{tabular}
\end{table}

\subsection{Surface coating and radon emanation}\label{sec:AdditionalSurfaceTreatments}

In addition to polishing methods, other surface treatments were tested to assess their impact on radon emanation rates. These treatments included applying epoxy coating and covering with aluminized Mylar. The results of these treatments are summarized in Table \ref{tab:SurfaceTreatments} and illustrated in the accompanying images.

\begin{figure}[!th]
\begin{center}
\includegraphics[height=6cm]{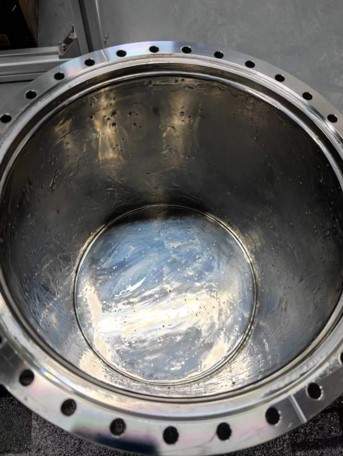}
\includegraphics[height=4.5cm, angle=90]{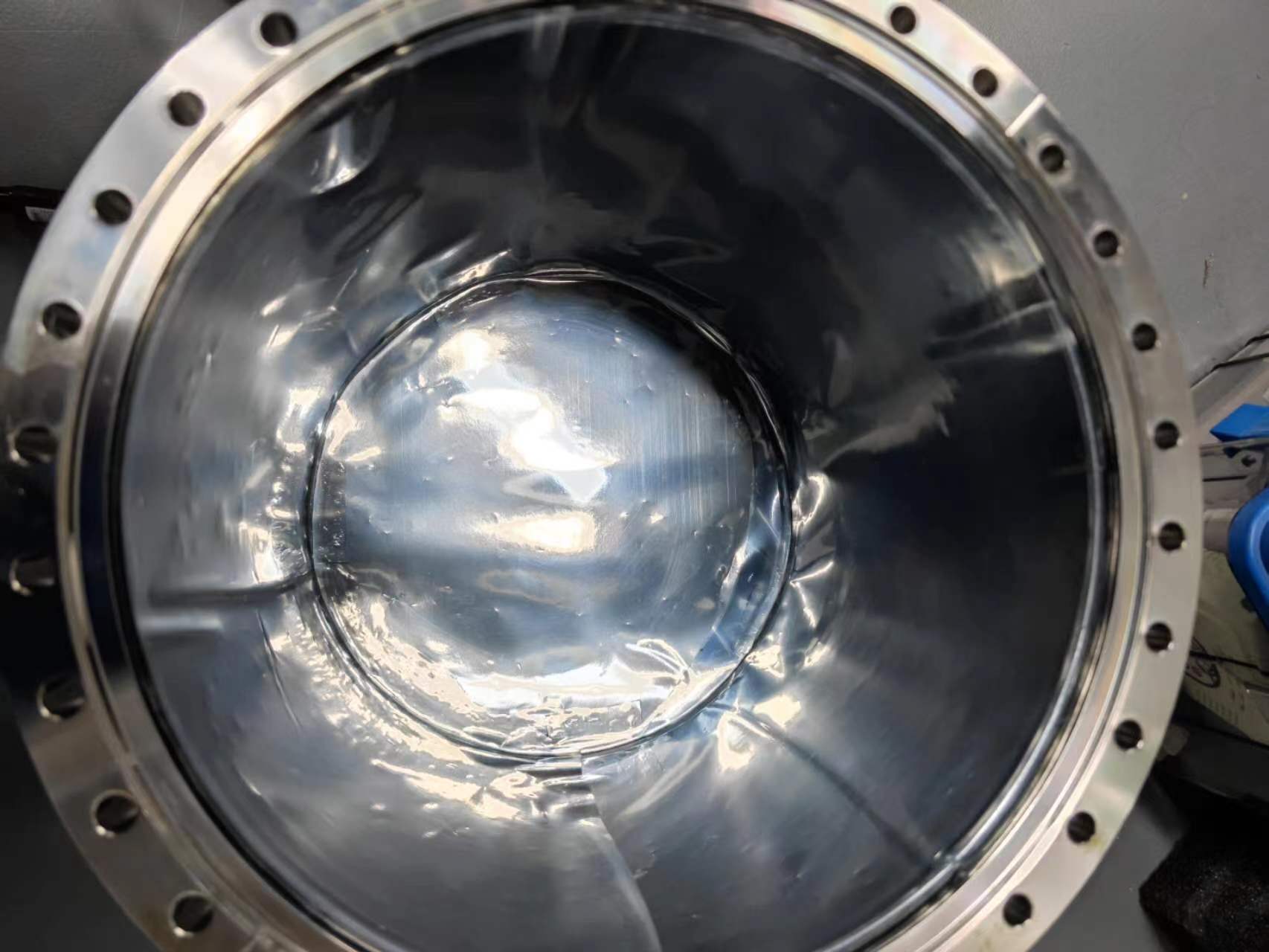}
\caption{Stainless steel chambers with surface coating. (Left) Epoxy coating. (Right) Mylar membrane. }
\label{fig:treatments}
\end{center}
\end{figure}

\begin{table}[h!]
\centering
\caption{Effect of surface coating on radon blanks.}
\label{tab:SurfaceTreatments}
\small
\begin{tabular}{|c|c|c|}
\hline
\textbf{Surface Treatment} & \textbf{Blank [mBq/m$^2$]} & \textbf{Reduction Factor} \\
\hline
Chamber 1 (untreated) & 6.94 ± 1.25 & \multirow{2}{*}{(90 ± 7)\%} \\
\cline{1-2}
Chamber 1 + Epoxy & 0.71 ± 0.44 & \\
\hline
Chamber 2 (untreated) & 5.04 ± 0.98 & \multirow{2}{*}{(60 ± 12)\%} \\
\cline{1-2}
Chamber 2 + Mylar & 2.03 ± 0.47 & \\
\hline
\end{tabular}
\end{table}

In the epoxy treatment, the inner surface of the stainless steel chamber including its body and upper flange was brushed with epoxy (LOCTITE\textsuperscript{\textregistered}) \cite{Henkel}, followed by a drying period of one day. A second coat of epoxy was then applied to ensure uniform coverage, and the chamber was left to dry for another day. This treatment resulted in a significant reduction in radon emanation, achieving a reduction factor of ($90 \pm 7)\%$ compared to untreated results.

For the aluminized Mylar treatment, a double-layer Mylar membrane was cut to the appropriate size, and apply solid glue to one side. The membrane was then adhered to the inner surface of the chamber. This treatment achieved a reduction in radon emanation of ($60 \pm 12)\%$ compared to untreated results.

\section{Conclusion}\label{sec:Conclusion}

This study details the development and evaluation of two advanced radon measurement systems designed to screen radon emanation rates from components of rare event search experiments, with the goal of replacing or improving those with high radon emanation to meet the stringent low-radon requirements. The radon emanation measurement system is ideal for small sample measurements, while the radon trap system, achieving about 30 times increase of sensitivity, is suited for assessing radon emanation from larger containers.

Our experimental findings demonstrate a strong correlation between material surface roughness and radon release, with smoother surfaces leading to lower radon emanation rates. Polishing methods, including electrochemical and mirror polishing, were effective in reducing surface roughness and, consequently, radon release.

Further mitigation techniques, such as epoxy coating and Mylar covering, showed substantial reductions in radon emanation rates. These treatments highlight practical approaches to controlling radon-induced backgrounds in low-background experiments.

In summary, the radon measurement systems and surface treatment methods developed in this study provide effective solutions for controlling radon levels in rare event search experiments. Implementing these techniques will enhance the sensitivity of rare event searches, including dark matter detection and NLDBD search.

\section*{Acknowledgement}

This work has been supported by the Ministry of Science and Technology of China (No. 2023YFA1606204), the National Natural Science Foundation of China (No. 12222505).

The authors would like to express their gratitude to Shoukang Qiu from Nanhua University for providing the $^{226}$Ra source used in the calibration experiments.

\bibliography{mybibfile}
\end{document}